\newcommand{\be}{\begin{eqnarray}}
\newcommand{\ee}{\end{eqnarray}}
\newcommand{\nn}{\nonumber}
\newcommand{\p}[1]{(\ref{#1})}
\newcommand{\lb}[1]{\label{#1}}
\newcommand\s{\scriptscriptstyle}

\newcommand\q{\quad}

\newcommand\Tr{\mbox{Tr}\,}

\def\g{\gamma}

\def\d{\delta}

\def\ve{\varepsilon}

 \def\th{\theta}

\def\pa{\partial}

\newcommand\A{{\s A}}

\newcommand{\pp}{{++}}
\newcommand{\m}{{--}}

\newcommand{\Dp}{D^{\pp}}
\newcommand{\Dm}{D^{\m}}

\newcommand{\Vp}{V^\pp}
\newcommand{\Vm}{V^\m}

\def\sfrac#1#2{{\textstyle\frac{#1}{#2}}}

\documentclass[11pt]{article}

\usepackage{amsmath,amssymb}
\usepackage{graphicx}
\usepackage{cite}
\usepackage{bm}

\def\theequation{\arabic{section}.\arabic{equation}}
\begin{document}
\renewcommand{\thefootnote}{\fnsymbol{footnote}}

\vskip 15mm

\begin{center}

{\Large Conformal properties of hypermultiplet actions in six dimensions}

\vskip 4ex

E.A. \textsc{Ivanov}$\,^{1}$,
A.V. \textsc{Smilga}\,$^{2}$,

\vskip 3ex

$^{1}\,$\textit{ Bogoliubov  Laboratory of Theoretical Physics, JINR, 141980 Dubna,
Moscow Region, Russia}
\\
\texttt{eivanov@theor.jinr.ru},
\\[3ex]
$^{2}\,$\textit{SUBATECH, Universit\'e de
Nantes,  4 rue Alfred Kastler, BP 20722, Nantes  44307, France
\footnote{On leave of absence from ITEP, Moscow, Russia.}}
\\
\texttt{smilga@subatech.in2p3.fr}
\end{center}

\vskip 5ex

\begin{abstract}
\noindent We 
consider scale-invariant interactions of 
 $6D$   ${\cal N}{=}1$  hypermultiplets with the gauge multiplet. 
If the canonical dimension of the matter scalar field is assumed to be
1, scale-invariant lagrangians involve higher 
derivatives in the action. Though scale-invariant, all such lagrangians are {\it not} invariant with respect
to special conformal transformations and their superpartners.
If the scalar canonical dimension  is assumed to be 2, conformal invariance holds at the classical, but not at the 
quantum
level.
\end{abstract}

\renewcommand{\thefootnote}{\arabic{footnote}}
\setcounter{footnote}0
\setcounter{page}{1}
\section{Introduction}
For the last 25 years, supersymmetric gauge theories in higher dimensions have been 
a subject of intense studies. A special attention was focused on $10D$ SYM theories 
due to their close relation to superstrings. Since an efficient off-shell $10D$ superfield 
approach is lacking, the only possibility to formulate and study  these 
theories is in a component formalism. The main difficulty hindering the superspace 
description consists in a large number of odd coordinates ($10D$ Majorana-Weyl spinor has 16 real components). An unconstrained superfield
involves too many ($\approx 2^{16}$) component fields. So, to write down a meaningful action, one has to impose the appropriate constraints, 
and in $10D$ we do not know how to accomplish this. An alternative 
is to make use of the lower-dimensional superspace formulations (see e.g. \cite{LD}), but then only a part 
of the original supersymmetry  is manifest.   

In six dimensions, the superspace of minimal Grassmann dimension involves only 8 odd coordinates 
which are comprised in the Weyl $(1,0)$ $6D$ spinor $\theta^a_i$ subject to the pseudoreality
condition
\be
\lb{pseudoreal}
{\th^a_i}\ = \  - \epsilon_{ij} C_{{b}}^a(\th^b_j)^*  \,
\ee
(with $C$ being an antisymmetric charge conjugation matrix).  As a result, 
an off-shell  superfield description of ${\cal N}=(1,0)$ $6D$ SYM theories 
(as well as of matter hypermultiplets) exists \cite{6D}.\footnote{ 
One can equally well consider the ${\cal N}=(0,1)$ $6D$ superspace and superfield theories in it.} 
On the basis of the harmonic superspace (HSS) approach \cite{HSS,HSSbook} it was worked 
out in \cite{Zupnik} (see also \cite{UK}). The standard off-shell ${\cal N} =1$ $6D$ SYM action is 
constructed in the HSS framework in the same way as in the case of the extended  ${\cal N} =2$ $4D$ SYM  
theory \cite{Zupnik2,HSSbook}. It has canonical dimension 4. In $4D$, this implies a dimensionless coupling
constant, scale and conformal invariance of the action (at the classical and, for  ${\cal N} =4$ theories, also quantum levels). 
But its 6--dimensional counterpart involves 
a dimensionful coupling constant and for this reason is neither
 conformal nor  renormalizable.

In Ref.\cite{ISZ} we constructed a nontrivial $6D$ SYM lagrangian with canonical dimension 6. Being expressed in components, 
the corresponding action looks as
   \be
 S &=& -\frac{1}{g^2}\int d^6x  \, \mbox{Tr}
\left\{ \left( \nabla^M F_{ML}\right)^2  +  i\psi^j\gamma^M \nabla_M (\nabla)^2\psi_j
 + \frac 12 \left(\nabla_M{\cal D}_{jk}\right)^2
\right. \nn \\
&& \q \left.
+\,  {\cal D}_{lk}{\cal D}^{kj}{\cal D}^{\;\;\;l}_{j}
  -2i {\cal D}_{jk} \left( \psi^j\gamma^M\nabla_M\psi^k
- \nabla_M \psi^j \gamma^M \psi^k \right)
+ (\psi^j\gamma_M \psi_j)^2  \right. \nn \\
&& \q \left. +\, \frac 12  \nabla_M\psi^i
\gamma^M\sigma^{NS}[F_{NS}, \psi_j]
- 2\nabla^M F_{MN}\, \psi^j\gamma^N\psi_j
 \right\}\, ,
\label{CompAct}
 \ee
where $M,N,\ldots = 0,1,\ldots,5$ are Lorentz vector indices, the extra ``flavor'' indices $j,k,l = 1,2$ are raised and
lowered with the antsymmetric 
tensors $\epsilon_{jk}$, $\epsilon^{jk}$ ($\epsilon_{12} = \epsilon^{21} =1, 
\epsilon^{kj}\epsilon_{jl} = \delta^k_l$), $F_{MN}$ is the gauge field strength, $\nabla_M$ is the covariant derivative, 
$\psi^k$ are the fermion
spinor  fields satisfying the constraint (\ref{pseudoreal}), and $D^{jk}=D^{kj}$ are scalar fields of canonical dimension 2 
(they are auxiliary in the case of the standard action, but are propagating in the 
action \p{CompAct}).  The action \p{CompAct} involves a dimensionless coupling $g$ and is scale-invariant. One can show that at the classical 
level it is also conformally invariant and 
in fact respects the whole $6D$ superconformal group $OSp(8^\star|2)$. A dimensionless coupling 
constant means
renormalizability, which allows one to calculate loops. In \cite{ISZ}, we calculated the $\beta$ function of this theory 
and found
that it is non-zero and positive, corresponding to the zero charge situation (like in the ordinary $4D$ QED). This breaks conformal 
(and superconformal) invariance in quantum theory.

The price one has to pay for renormalizability is the presence of higher derivatives (HD) in the lagrangian. Usually, HD
theories are plagued by ghosts. Physically, this means that the spectrum of a theory is bottomless: 
it contains a continuum set of states
going all the way to $E = -\infty$ (cf. the phenomenon of falling into the center in the ordinary quantum mechanics \cite{fall}).
Presumably, the theory (\ref{CompAct}) also suffers from this sickness. We argued in Refs.\cite{TOE} 
that a $6D$ theory where 
conformal symmetry
is {\it exact} (does not involve a quantum anomaly) might be benign in this respect and the vacuum ground state could be well defined
there. An attractive possibility is that this theory enjoys the maximum ${\cal N} =(2,0)$ $6D$ superconformal symmetry. 
Unfortunately, a conventional field theory realization of this symmetry at present is not known.      
 Actually, the {\it only} known today nontrivial (i.e. involving nontrivial interactions) 
$6D$ action invariant under a linear realization of superconformal 
group and involving fields with canonical scale dimension 
 (1 for bosons, 3/2 for fermions) is the action (\ref{CompAct}).\footnote{The gauge-fixed brane $6D$ actions respecting invariance under 
field-dependent nonlinear realizations of the $6D$ $(2,0)$ superconformal group as the group of motion of the proper super AdS backgrounds
 were considered in \cite{Nonl}.}     

In this paper, we perform a search for some other superconformal actions (with the motivation to find eventually an anomaly-free superconformal theory). 
Our basic idea is to include the interactions of the gauge superfield
with the matter superfields (we know that in four dimensions this {\it is} the way to construct truly conformal actions).
The simplest $6D$ matter supermultiplet is a hypermultiplet 
(after dimensional reduction, it gives ${\cal N} =2$ $4D$ hypermultiplet). We  tried to build up  
 a conformally invariant lagrangian involving such a multiplet (a spinor hypermultiplet is also briefly discussed). 
Let us say right away that we {\it failed} to do this. 
Still the reasons for such a failure seem to be interesting enough to make them a public knowledge.
 
In the next Section we start with reminding some basic definitions of $6D$ harmonic superspace and superconformal 
group. In Sect. 3,  we construct  scale-invariant lagrangians involving {\it (i)}
the  hypermultiplet $q^+$  of canonical 
dimension 1 
and {\it (ii)} the spinor Grassmann multiplet $F^{+a}$ of canonical dimension $3/2$ (in both cases, the fermion components have
dimension $3/2$). We show that these lagrangians are {\it not} invariant, however,
 with respect to special conformal transformations. 
In Sect. 4 we consider the hypermultiplet of ``unnatural'' canonical dimension 2. Here, it is easy to construct a conformally 
invariant
classical action, but, like for the action (\ref{CompAct}), the conformal symmetry
 is broken by quantum effects. The sign of the $\beta$ function
is the same as for the pure gauge theory and corresponds to the zero charge situation. In addition, quadratically divergent
counterterms appear.

\section{Harmonic  $6D$ superspace and matter superfield actions}
 \setcounter{equation}0
We start with a brief account of some basic definitions, addressing the reader 
to Ref.\cite{ISZ} for details. 

The algebra of the minimal ${\cal N}=(1,0)$
$6D$ supersymmetry 
enjoys the automorphism (or $R$-symmetry) $SU(2)$. 
We introduce the coset   $SU(2)/U(1) \sim S^2$  parametrized by the  harmonics 
$u^{\pm j}$ ($u^-_j = (u^{+j})^*$,
$u^{+j}u^-_j = 1$). Introducing the projections $\theta^{\pm a}= u^\pm_j \theta^{ja}$,  
we can define the $6D, \;{\cal N}{=}1$ harmonic  superspace (HSS) as a superspace 
parametrized by the coordinates
$(x^M, \theta^{\pm a}, u^{\pm i})$. The {\it analytic} superspace is a subspace of 
$6D$ HSS parametrized by the coordinates $(x^M_A, \theta^{+ a}, u^{\pm i})$ where 
$$x^M_\A=x^M+ \frac i2 \th^{+a}\g^M_{ab}\th^{-b}\, .$$ 
In the expression above, $\gamma_{ab}$ are antisymmetric
Weyl $6D$ matrices satisfying the condition
  \be
&&(\gamma_M)_{ac}(\tilde\gamma_N)^{cb}+(\gamma_N)_{ac}(\tilde\gamma_M)^{cb}
=-2\d^b_a
\eta_{MN}\,
\ee
with $\tilde\gamma_M^{ab}=\sfrac12\ve^{abcd}
(\gamma_M)_{cd}$. The characteristic feature of the analytic superspace  
$(x^M_A, \theta^{+ a}, u^{\pm i})$ is that it is closed under ${\cal N}=(1,0)$ Poincar\'e 
supersymmetry transformations (as well as under a wider superconformal ${\cal N}=(1,0)$ group).

One can further introduce covariant spinor $D^{\pm}_a$ and  harmonic $D^{\pm\pm}$ derivatives. 
In the central basis $(x^M, \th^\pm, u)$, they have the following form
 \be
D^+_a  =  \frac {\partial }{\partial \theta^{-a}} - \frac i2 
(\gamma^M \theta^+)_a \frac 
{\partial}{\partial x^M}\,, \quad 
  D^-_a = -\frac {\partial }{\partial \theta^{+a}} - \frac i2  
(\gamma^M \theta^-)_a \frac 
{\partial}{\partial x^M} \label{CovD}
 \ee
and 
 \be
 D^{++} =  u^{+i} \frac {\pa }{ \pa u^{-i}} \equiv \partial^{++} , \q 
D^{--} =  u^{-i} \frac {\pa }{ \pa u^{+i}} \equiv \partial^{--}\ .
  \ee
In the analytic basis $(x^M_A, \th^\pm, u)$ the derivative $D^+_a$ becomes short while $D^{\pm\pm}$ 
get additional terms
\be
D^+_a = \frac {\partial }{\partial \theta^{-a}}\,, 
\ee
\be
D^{\pm\pm} = \partial^{\pm\pm} + \frac{i}{2}\theta^{\pm}\gamma^M\theta^{\pm} \frac{\partial}{\partial 
x^M_A} 
+\theta^{\pm a}\frac{\partial}{\partial \theta^{\mp a}}\,. \label{DHanal} 
\ee  
The commutator of harmonic derivatives $D^{++}$ and $D^{--}$ yields the operator $D^0$ which 
counts the external ``harmonic'' $U(1)$ charges of harmonic superfields (the charges with respect to 
$U(1)$ in the denominator of the harmonic 2-sphere $SU(2)/U(1)$)
\be
[D^{++}, D^{--}] = D^0\,. \label{D0} 
\ee
In the central and analytic bases this operator is given, respectively,  by
\be
&&\underline{\mbox{central basis:}} \quad D^0 = u^{+i}\frac{\pa}{ \pa u^{+i}} 
- u^{-i} \frac {\pa}{ \pa u^{-i}}\,, \nn \\ 
&& \underline{\mbox{analytic basis:}} \quad D^0 = u^{+ i}\frac {\pa}{ \pa u^{+i}} - u^{-i} \frac {\pa}{ \pa u^{-i}} + 
\theta^{+a}\frac{\partial}{\partial \theta^{+ a}} - \theta^{-a}\frac{\partial}{\partial \theta^{- a}} \,. 
\ee
The measures of integration over the full $6D$ HSS and its analytic subspace 
$\mu$ $\equiv$ $d^6x du d^4\theta^+ d^4\theta^-$ and $d\zeta^{(-4)} du \equiv 
d^6x du d^4\theta^+$ (where $du$ stands for integration over harmonics) are related as 
\be
d^6 x du d^4\theta^+ d^4\theta^- = d^6 x_A du d^4\theta^+ (D^+)^4\,, \quad (D^+)^4 \equiv 
-\frac{1}{24}\varepsilon^{abcd}D^+_a D^+_b D^+_c D^+_d \,.\label{RelMea}
\ee
For brevity, hereafter we will frequently omit the index ``A'' of the bosonic coordinate 
of the analytic superspace.

The basic ingredient for constructing a gauge-invariant
 lagrangian is the analytic gauge superfield
$V^{++}$. It is an anti-hermitian matrix which belongs to the Lie algebra of the corresponding gauge group,
depends only on $(x^M_A, \th^+, u)$ in the analytic basis and has the charge 2 with respect to the operator $D^0$:
 \be
(\mbox{a}) \;\; D^{+}_a V^{++} = 0\, \qquad (\mbox{b}) \;\; D^{0} V^{++} = 2 V^{++}\,. \label{DefV++}
 \ee
In the analytic basis the condition $(\mbox a)$, in view of ``shortness'' of $D^{+}_a$, simply 
states that $V^{++}$ does not depend on $\theta^{-a}\,$. The relations \p{DefV++} 
can be regarded as a basis-independent definition of $V^{++}$. 

We need also the non-analytic gauge superfield $V^{--}$ which is related to $V^{++}$ through 
the ``zero harmonic curvature'' condition
   \be
  \lb{A2}
  \Dp\Vm-\Dm\Vp+[\Vp,\Vm]=0\,.\lb{hzc}
   \ee   
 
The component off-shell action (\ref{CompAct}) is obtained after fixing the Wess-Zumino 
gauge in the superfield action:
  \be
\label{dejstvie}
S = \frac{1}{2g^2}\int  d^6x du d^4 \th^+ \, \Tr\left(F^\pp\right)^2 \lb{hactan}\ ,
\ee
 where $F^\pp = (D^+)^4 V^{--}$ is an analytic superfield, and doing there the Berezin and 
harmonic integrals. One can check that this action is invariant with respect to the supergauge transformations  
 \be
\lb{supgauge}
\delta V^{++} = D^{++}\Lambda +[V^{++}, \Lambda] \ ,
\ee
 where $\Lambda$ is an analytic superfield
of zero harmonic charge. This gauge freedom allows one to eliminate an infinite number of component
 fields which appear 
after expanding $V^{++}(x,\theta, u)$ in $\theta$ {\it and} in harmonics. In a Wess-Zumino 
gauge, $V^{++}$ acquires a simple compact form
  \be
\label{VppWZ}
V^{++}_{(WZ)} = \frac 12 \theta^{+}\gamma^M \theta^{+} A_M +
\frac {\sqrt{2}}3 \epsilon_{abcd}\theta^{+a}  \theta^{+b} \theta^{+c} \psi^{-d} +
\frac 18   \epsilon_{abcd}\theta^{+a}  \theta^{+b} \theta^{+c} \theta^{+d}  {\cal D}^{--} \label{WZg} \, ,
\ee 
where $\psi^{-a} = \psi^{aj}u^-_j\,, \quad {\cal D}^{--} = {\cal D}^{jk}u^-_ju^-_k$ and the fields 
$A_M, \psi^{aj}, D^{jk}$
depend only on $x^M\,$. The only residual gauge freedom is with respect to the ordinary gauge transformations, with $A_M$ transforming 
as the standard Yang-Mills field.  

To describe matter fields, we consider a pair of hypermultiplet analytic superfields $q^{+A}$ 
and write the invariant action as (this action has the same form both in $4D$ and $6D$ HSS)
 \be
\label{q+act}
 S \ =\ -\frac 12 {\rm Tr} \int d^6x_Adu d^4\theta^+  \,  q^{+A} \nabla^{++} q^+_A \ .
 \ee
Here $\nabla^{++} = D^{++} + V^{++} $ is the covariantized harmonic derivative and $A= 1,2$ is a subflavor (or Pauli-G\"ursey) 
doublet index lowered and raised by $\epsilon_{AB}, \epsilon^{AB}$. 
If the superfield $q^{+A}$ is placed into a real representation of the gauge group, it can be subject 
to the reality condition $\widetilde{q^{+ A}} = \epsilon_{AB}q^{+ B}$ with respect to  
a properly defined (pseudo)conjugation $\widetilde{\;\;\;}$ (see \cite{HSSbook} for details). 
In what follows we shall deal with the case when $q^{+A}$ belongs to the adjoint representation, i.e. is represented by an anti-hermitean matrix 
like gauge superfields $V^{++}, V^{--}\,$. In this case, 
$\nabla^{++} = D^{++} + [V^{++}, \cdot]\,$.\footnote{We could equally consider any other representation. For complex representations, 
the extra $SU(2)_{PG}$ invariance would be broken down to $U(1)_{PG}$ unless one admits a doubling of fields. In fact,
 in the most part of the paper,
non-Abelian nature of the gauge group and of $q^{+A}$ will not be important.} 

The action (\ref{q+act}) is invariant  with respect to the gauge
transformations (\ref{supgauge}) supplemented by $\delta_\Lambda q^+ = [q^+, \Lambda]$. 
After expansion of $q^+$ in $\theta^+$ and $u$, we obtain
an infinite set of fields. In contrast to the case of gauge supermultiplet, where all such fields, besides those surviving in WZ gauge \p{VppWZ},
 are gauge degrees of freedom, in the hypermultiplet case the extra fields are auxiliary: they vanish or can be expressed in terms of physical fields 
on mass shell from equation of motion $\nabla^{++} q^{+A} = 0$. The situation is exactly the same as in four dimensions and we address the reader
 to Chapter 5 of the book \cite{HSSbook} for detailed explanations.\footnote{A description of the $6D$ hypermultiplet coupled to SYM  in 
$4D$ ${\cal N}=1$ superspace in the spirit of \cite{LD} (and also in the projective 
$6D$ ${\cal N}=1$ superspace) was given in a recent preprint \cite{GPT}.}   

As a result, we are left with a quartet of scalars $f_{jA}$ ($f_{j1}$ and $f_{j2}$ being related to each other
by the conjugation $\widetilde{\;\;\;}$) and a pseudoreal fermion $\psi^a_j$. The 
component lagrangian following
from (\ref{q+act}) has a standard form --- it is quadratic in (covariant) derivatives for bosons and is of the first order in derivative for fermions.  

If ascribing the usual canonical dimension 1 to scalar fields and the dimension $3/2$ to fermions, the lagrangian
  \be
\label{Lfpsi}
{\cal L} \sim (\partial f)^2 + i \psi /\!\!\!\! \partial \psi
  \ee
 has canonical dimension four and it is not scale-invariant in 
six dimensions. Let us try to write a scale-invariant $6D$ action and then explore its conformal properties. There are 
three distinct
ways to do this.
 \begin{enumerate} 
\item
We may stay with the standard 
hypermultiplet $q^+$, but write an action involving on top of $\nabla^{++}$ also 
the covariant derivative $\nabla^{--} = D^{--} + [V^{--}, \cdot ]\,$. 
The  simplest such action is
 \be
\label{qnab--q}
 S_1 =\  {\rm Tr} \int d^6x du d^4\theta^+ d^4\theta^-   \, q^{+A} \nabla^{--} q^+_A \equiv
\int d^6x du d^4\theta^+ d^4\theta^- {\cal L}_1\ ,
 \ee
but any structure 
  \be
\label{Sn}
 S_n =\  {\rm Tr} \int d^6x d^4\theta^+ d^4\theta^-  du \,  q^{+A} (\nabla^{--})^{n} (\nabla^{++})^{n-1}     
q^+_A \equiv \int d^6x du d^4\theta^+ d^4\theta^- {\cal L}_n 
 \ee
is possible.
 Note that the operator $\nabla^{--}$ brings about a nontrivial dependence on $\theta^-$ and we cannot write the action
by only doing the integral over $d^4\theta^+$. One has to integrate over all thetas. The extra $\int d^4\theta^-$ brings
 the dimension
$m^2$ and makes the actions (\ref{Sn}) dimensionless and the corresponding theories scale-invariant. The presense of an 
infinite set of lagrangians
with the same scaling dimension is a rather remarkable feature specific  to HSS. The point is that, in contrast to
ordinary $x$ or $\theta$ derivatives, harmonic derivatives $D^{\pm\pm}$ carry no dimension. The set (\ref{Sn}) forms a complete basis of scale-invariant 
actions. 
Any other candidate action, e.g. Tr $ \int d^6x d^8\theta  du 
q^{+A} \nabla^{--} (\nabla^{++})^2   \nabla^{--}   q^+_A$, can be re-expressed in terms of $S_n$  by repeatedly 
 using the commutation relation 
\be
[\nabla^{++}, \nabla^{--}] = [D^{++}, D^{--}] = D^0  \label{D00}
\ee
(the reason why covariantized derivatives satisfy the same commutation relation as the flat 
ones is the harmonic zero-curvature condition \p{hzc}). 

Unfortunately, it is not easy to express the actions (\ref{Sn}) in components. 
More precisely, one could do this,
 but such an action would involve
an infinite number of {\it dynamical} fields. The point is that higher components in the harmonic expansion of $q^+$ which in the standard 
lagrangian \p{q+act} were eliminated on mass shell due to equations of motion and had the status of auxiliary fields, are {\it not} 
eliminated anymore while considering the above 
higher-derivative actions. Indeed, 
by dimensional reasoning, the equations of motion
involve two extra derivatives, and an equation which in the case of the action (\ref{q+act}) looked as, 
say, $F=0\,$,  acquires now the form $\Box F = 0$ 
bringing about a propagating degree of freedom.
This is similar to what happened to the former auxiliary fields $D^{jk}$ of the gauge multiplet 
in the HD action (\ref{CompAct}). But in the case at hand we have an infinite set of such fields. 

\item We may ascribe ``unnatural'' dimensions 2 and $5/2$ to the scalar and fermionic components 
of $q^+\,$, respectively.
In this case, the lagrangian (\ref{Lfpsi}) has dimension 6 and the action is scale-invariant. 
It is easy to show that this invariance extends to the full superconformal 
invariance.\footnote{By ascribing appropriate canonical dimensions, one can write superconformal  hypermultiplet actions 
 with the standard numbers of derivatives  (two for bosons and one for fermions)  in any dimension $1 \leq D \leq 6$ 
following the  $4D$, ${\cal N}=2$ pattern \cite{HYP} (see also \cite{KI} in the harmonic superspace context).
Generically, these actions  involve non-trivial self-interactions.}

\item Finally, instead of dealing with the standard bosonic hypermultiplet 
superfield $q^{+A}$ we may consider an anticommuting
  spinor  analytic superfield $\Psi^{+aA}(x^M, \theta^+, u)$ with canonical dimension $3/2$. 
If it belongs to the adjoint color representation, its lowest component has the same quantum numbers as the fermion component
of the gauge supermultiplet. Then we could hope that, considering a higher-derivative 
 theory with both $\Psi^{+aA}$ and $V^{++}\,$, we could eventually find out a field-theoretic formulation of the mysterious 
$6D$, \ ${\cal N} = (2,0)$ superconformal theory  \cite{TOE}.
A natural way to write a scale-invariant lagrangian of $\Psi^{+aA}$ would be
  \be
\label{totder}
 S_\Psi \ =\  {\rm Tr} \int d^6x_A du d^4\theta^+  \, \Psi^{+aA} \nabla^{++} \nabla_M \gamma^M_{ab} \Psi^{+b}_A 
 \ee
  ($\nabla^M = \partial_M + [\Gamma_M, \cdot]\,$, where $\Gamma_M = 
\frac i4 (\tilde \gamma^M)^{ab} D^+_a D^+_b V^{--} $ is the superconnection). 
However, integrating by parts, one can show that 
the integral (\ref{totder}) vanishes identically.
One can  write a  non-vanishing action at cost of discarding not too much relevant 
Pauli-G\"ursey SU(2) symmetry. The corresponding non-vanishing 
action (with only $U(1)_{PG}$ symmetry) reads 
   \be
\label{PsiAct}
 S_\Psi \ =\  {\rm Tr} \int d^6x_A du d^4\theta^+   \, \Psi^{+aA} \nabla^{++} \nabla_M \gamma^M_{ab} 
\Psi^{+bB} 
(\tau^1)_{AB}\, .
 \ee
It is difficult to express it in components for the same reason as for the action (\ref{Sn}).
The superfield $\Psi^{+aA}$ involves an infinite number of components with growing isospins 
which survive on mass shell as dynamical degrees of freedom.
 \end{enumerate}

The presence of an infinite tower of propagating fields is a signal that 
the theories (\ref{Sn}) and (\ref{PsiAct}) are probably not feasible.
The second disappointing circumstance is that these actions are {\it not} conformally invariant. 
We show this in the next section.

 \section{Conformal properties of the HD actions}
\setcounter{equation}0
The ${\cal N}=(1,0)$, $6D$ superconformal group $OSp(8^\star|2)$ involves translations, dilatations, Lorentz rotations, special conformal transformations 
and also the transformations
of Poincar\'e and special conformal supersymmetries. To this set of space-time 
transformations one should also add $SU(2)$ transformations corresponding to the 
automorphism of the SUSY algebra. This $SU(2)$ appears in the anticommutator of the 
Poincar\'e and conformal supersymmetries and it acts non-trivially on $\theta^a_j$ and harmonics. 
General transformations were written in \cite{ISZ}. Since actually all other superconformal transformations are also contained 
in the closure of Poincar\'e and conformal supersymmetries, 
 it will be sufficient for our purposes to know only the transformations with respect to the latter. 
They are characterized by the Grassmann parameters 
$\eta^j_a$ (note that $\eta^j_a$  is the $(0,1)$ $6D$ spinor in contrast to $\theta^{ja}$ and the parameters of 
Poincar\'e supersymmetry $\epsilon^{ja}$ which are $(1,0)$ spinors),
\be
\delta_\eta u^+_k&=&2i\lambda^\pp u^-_k~,\q\delta_\eta u^-_k=0\,,\nn \\
\lambda^\pp &=& 2i\eta^+_a\th^{+a}\,, \; 
\Dm\lambda^\pp = 2i(\eta^-_a\th^{+a}+ \eta^+_a\th^{-a}) \equiv 2i\lambda\,, 
\label{Sconfharm} \\
\delta_\eta\theta^{+a}&=& i\eta^-_b\th^{+b}\th^{+a} -x^{ab}_\A\eta^+_b\,, \nn \\
\delta_\eta\theta^{-a} &=& i\eta^-_c\th^{-c}\th^{+a}-i\eta^+_c\th^{-c}\th^{-a} 
-\,i\eta^-_c\th^{+c}
\th^{-a} -\, x^{ac}_\A\eta^-_c\,, \label{SconfGrass}\\
\delta_\eta x^{ab}_\A&=& i\eta^-_cx^{bc}_\A\th^{+a} -\,i\eta^-_cx^{ac}_\A\th^{+b}\,,\label{SconfX}
\ee 
where $x^{ac} = \frac12(\tilde{\gamma}_M)^{ac} x^M\,$ and $\eta^\pm_a = \eta^i_a u^{\pm}_i\,$. 
We see that the analytic subspace $(x_A, \theta^+, u)$ is invariant under these transformations.
 The corresponding transformations 
of harmonic derivatives $D^{\pm\pm}$ and the gauge superpotentials $V^{\pm\pm}$ are 
\be
\delta_\eta D^{++} &=& -2i\lambda^{++}D^0\,, \quad \delta_\eta D^{--} = -2i \lambda D^{--}\,, \nn \\
\delta_\eta V^{++} &=& 0\,, \quad \delta_\eta V^{--} = -2i \lambda V^{--}\,, \label{SconfV}
\ee 
whence
\be
\delta_\eta \nabla^{++} = -2i\lambda^{++}D^0\,, \quad \delta_\eta \nabla^{--} 
= -2i \lambda \nabla^{--}\,. \label{NablaTran}
\ee

We will also need to know how the analytic superspace and full harmonic superspace integration 
measures are transformed. From \p{Sconfharm} - \p{SconfX} 
it follows that the analytic superspace integration measure is transformed as 
\be
\delta_\eta \,(d^6x_A du d^4 \theta^+) &=& \ d^6x_A du d^4 \theta^+  \left(\partial_{ab}\delta_\eta x^{ab} + 
2i\partial^{--}\lambda^{++} - \partial_{+ a}\delta_\eta\theta^{+a}\right) \nn \\ 
&=& 
-4i d^6x_A du d^4 \theta^+ \,(\eta^-_a \theta^{+a})\,   .
\ee
The full measure $\mu = d^6x du d^8\theta$ is transformed as 
\be
 \label{meastran}
 \delta_\eta\,\mu \ =\ \mu\,\left(\partial_{ab}\delta_\eta x^{ab} + 
2i\partial^{--}\lambda^{++} - \partial_{+ a}\delta_\eta\theta^{+a} 
- \partial_{- a}\delta_\eta\theta^{-a}\right) \ =\ i 
\mu\left(3\eta_a^+ \theta^{-a} + \eta_a^- \theta^{+a} \right). 
 \ee

To study how the actions (\ref{Sn}) and (\ref{PsiAct}) are changed under the above  
transformations, it would be sufficient to study free theories
with $V^{++} = V^{--} = \Gamma_M =0$ and $\nabla^{\pm\pm} = D^{\pm\pm},\ \ \nabla_M = \partial_M$, since 
the covariant derivatives $\nabla^{\pm\pm}$ transform in the same way 
as the non-covariantized ones. The actions of free theories are 
quadratic and
their color structure is trivial. To avoid confusion, we will continue to assume that the matter fields
belong to the adjoint representation and keep the color trace symbol, but actually  it is redundant
in what follows.
 
Let us discuss the spinor multiplet $F^{+aA}$ first, taking for simplicity 
only free part of the action \p{PsiAct}. As follows from the transformation laws 
\p{Sconfharm} - \p{SconfX}, the variation $\delta_\eta \partial_M$ involves the terms
with derivatives $\partial/\partial \theta^{+a}\,$. As a result, we obtain for the variation
 \be
\label{confvarspin}
 \delta_\eta S_{\Psi} \sim {\rm Tr} \int d^6x_A d^4 \theta^+ du \,  F^{+aA} \left(\eta^+_b \frac \partial {\partial \theta^{+a}} -
 \eta^+_a \frac \partial {\partial \theta^{+b}} \right) D^{++} F^{+bB} (\tau^1)_{AB} \nonumber \\
+ \ {\rm terms\  not\  involving} \  \theta \ {\rm   derivatives.}
 \ee
This does not vanish. We conclude 
that the action (\ref{PsiAct}) is not conformal.

Consider now the standard hypermultiplet action (\ref{q+act}). 
The term coming from $\delta_\eta\nabla^{++}$ in \p{q+act} is $\propto {\rm Tr} \epsilon^{AB} q^+_A q^+_B$ , which vanishes. 
Thus, the variation of the lagrangian in \p{q+act} is proportional to the  sum of the 
weight factors coming from the transformation of the measure and that of $q^{+ A}\,$. 
The scalar analytic superfield $q^{+ A}$ should be transformed with the same weight factor as the measure,
  \be
\label{qcftran}
\delta_\eta q^+ = i\mbox{d}\, (\eta^-_a \theta^{+a}) q^+ \ ,
 \ee
where $\mbox{d}$ is the scale dimension of $q^+$ equal to the scale dimension of
the lowest scalar component of $q^{+}\,$. The canonical choice is $\mbox{d}=1$, in which case the action (\ref{q+act}) is not conformally 
invariant in six dimensions (while its 4-dimensional 
counterpart {\it is}).

The six-dimensional action can be made conformally invariant for the non-canonical 
choice $\mbox{d}=2$. We will discuss such theory in some more details in the next section. 
Now let us turn to the HD actions (\ref{Sn}) with the standard choice $\mbox{d}=1$ for the 
scale dimension of the scalar field.

Consider first the action $S_1$. Its free part vanishes identically  due to 
analyticity of $q^{+a}$. This can be checked by trading $\int d^4\theta^-$ for  $(D^+)^4$  and using the commutation relations of $D^{+}_a$ 
with $D^{--}$ and $D^{-}_a$, 
which are easy to find looking at the explicit expressions \p{CovD} - \p{DHanal}. $S_1$ involves, however, nontrivial interactions
coming from the term $\sim V^{--}$ in $\nabla^{--}$ 
(see Appendix B). Since 
$\nabla^{\pm\pm}$ have the same transformation laws \p{NablaTran} as their flat counterparts, i.e., 
the  superconformal properties of $S_1$ are derived straightforwardly. 
One  obtains 
 \be
\label{delS1}
 \delta_\eta S_1 = i \int \mu\,\lambda\, {\cal L}_1\ .
 \ee
Consider now a generic action $S_n\,$. Its conformal variation is given by
  \be
\label{delSndot}
 \delta_\eta S_n &=& i \int \mu \left[
\left(3\eta^+\theta^- + \eta^- \theta^+ - 2n\lambda + 2\eta^- \theta^+ \right) {\cal L}_n + \ldots \right] 
\nn \\ 
&=& i \int\mu \left[ (3-2n)\,\lambda\, {\cal L}_n + \ldots \right],
 \ee
where the term $3\eta^+\theta^- + \eta^- \theta^+$ comes from the variation of the measure, 
the term $ -2n\lambda$ 
from  $\delta_\eta \nabla^{--}$, the term $2\eta^-\theta^+$ from  $\delta_\eta q^+$ and 
dots stand for the terms appearing when harmonic derivatives hit the weight factors 
(and $\lambda^{++}$ in $\delta_\eta \nabla^{++}$). 
A remarkable fact proved in  Appendix A is that all such terms are reduced to total harmonic derivatives and 
 do not contribute into the variation of the actions $S_n\,$. Thus, the result
  \be
\label{delSn}
 \delta_\eta S_n
= i(3-2n) \int\,\mu\,\lambda {\cal L}_n 
 \ee
 is exact ! 
\footnote{For $S_1$, the vanishing of the term generated while acting by $\nabla^{--}$  
on the factor $\eta^-\theta^+$ in  $\delta_\eta q^+$ (which produces $\lambda^{--} 
= \eta^-_a \theta^{- a}$) follows immediately from the property Tr $q^{+A} q^+_A = 0\,$.}

We conclude that neither the actions $S_n$ themselves, nor any linear combination of them are conformal.

\section{Conformal anomalies}
\setcounter{equation}0
As already mentioned, the action (\ref{q+act}) is conformally invariant at the classical level if ascribing unnatural scale dimension $\mbox{d}=2$ 
to $q^{+A}$ and assuming that $\delta_\eta q^{+A} = 2i(\eta^- \theta^+) q^{+A}\,$. 
In this section, we discuss the theory with the action (\ref{q+act})
supplemented by the HD action (\ref{CompAct}) for the gauge field. In Ref.\cite{ISZ} we showed that the theory (\ref{CompAct})
alone involves conformal anomaly. Let us study now the anomalies coming from the matter sector. The problem boils down to calculating
the loop of matter fields in the background of gauge multiplet and determining the corresponding correction to the gauge fields effective action. 
We perform
this calculation in components.

The lowest component of the superfield $q^{+A}$ can be written as $f^{jA}u_j^+$, where $f^{jA}$ is pseudoreal
$$ \overline{f^{jA}} = -f_{jA} $$
as a consequence of the property $\widetilde{q^{+ A}} = q^+_A$. 
The scale dimension of $f$ is 2. It is convenient to choose as  background the superfield (\ref{VppWZ}) involving only the auxiliary components
$\propto D^{jk}$. As is clear from (\ref{q+act}), the field $D^{jk}$ interacts only with the lowest scalar component of $q^{+A}$ and all 
other components are
irrelevant. The interaction lagrangian is
   \be
 \label{LffD}
{\cal L}_{int} =  - g  {\rm Tr} \{f^{jA} f^k_A D_{jk} \}\ ,
 \ee
 where we used the perturbative normalization $\nabla^{++} = D^{++} + g[V^{++}, \cdot]$. 
 To find the corrections to the effective gauge lagrangian, one has to evaluate the graphs in Fig. 1, the graph in Fig. 1a giving the
counterterms $\propto D^2$ and  $\propto (\partial D)^2$ and the graph in Fig. 1b giving 
the counterterms  $\propto D^3$. To perform the calculation, 
we need the scalar
propagator which follows from the corresponding kinetic term. The latter can be derived from (\ref{q+act}) and has the form
 \be
\label{Lff}
 {\cal L}_{kin} \ =\ - \frac 12  {\rm Tr} \{ \partial_M f_{jA}  \partial_M f^{jA} \}\ .
 \ee
This implies the propagator
 \be
\label{propf}
 \left \langle f_{jA}^a f_{kB}^b \right \rangle \ =\ -\frac {2i}{p^2} \delta^{ab} \epsilon_{jk} \epsilon_{AB}\ ,
 \ee
Here $a,b$ are color indices and $f = f^a t^a$. Consider first the graph in Fig. 1a. A straightforward calculation gives 
the following contribution to 
${\cal L}_{\rm eff}$,
 \be
\label{intD2}
\Delta {\cal L}_{\rm eff} \ =\ i {g^2c_V} {\rm Tr} \{ D_{jk} D^{jk} \} \int \frac {d^6p}{(2\pi)^6 p^2 (p+k)^2}\ .
 \ee
 We see that the integral involves a quadratic divergence. The term $\sim \Lambda_{UV}^2 D^2$ 
is the part of the {\it standard} SYM lagrangian in six dimensions with the canonical dimension 4. When evaluating \cite{ISZ} 
the coefficient of such a counterterm in the pure HD SYM theory in (\ref{CompAct}), we found it to vanish. But here it does not. 
\footnote{Cf. noncancellation of quadratically divergent scalar masses in usual  $\phi^4$ or Yukawa theories in four dimensions.}

\begin{figure}[h]
   \begin{center}
 \includegraphics[width=4.0in]{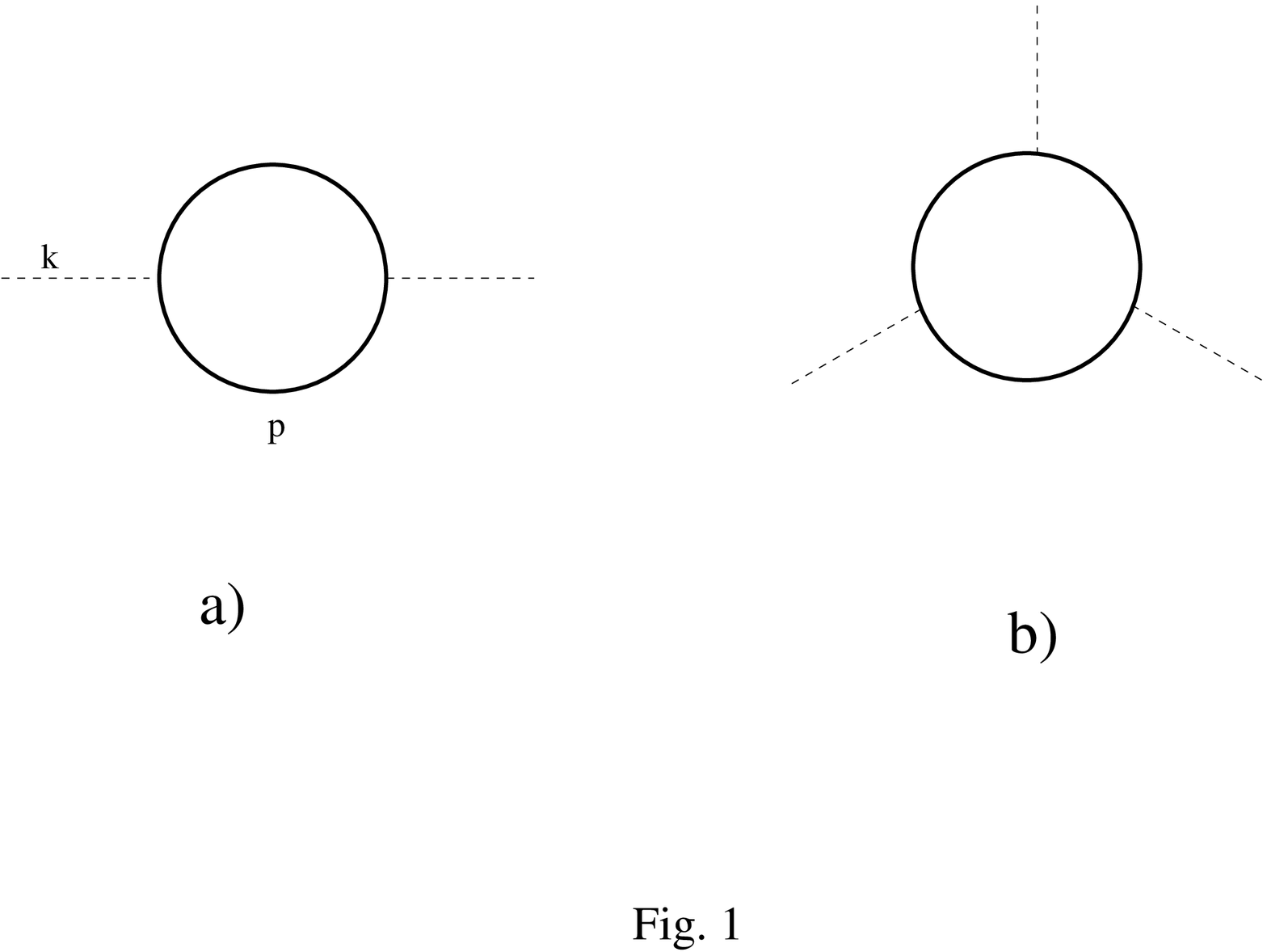}
        \vspace{-2mm}
    \end{center}
\end{figure}

Besides, there are counterterms  $\propto (\partial D)^2$  and  $\propto D^3$ of dimension 6. The former comes from the graph in Fig.1a. 
The coefficient
can be fixed if expanding the integral (\ref{intD2}) in external momentum $k$ (in the coordinate representation, this is the derivative expansion).
 We obtain after Wick rotation
  \be
\label{DLpD2}
\Delta {\cal L}_{\rm eff}^{(\partial D)^2} \ =\ - \frac {g^2c_VL}{3} {\rm Tr} \{ \partial_M D_{jk} \partial_M D^{jk} \}\ , 
 \ee
where
 \be
\label{L}
L = \int_\mu^\Lambda \frac {d^6p_E}{(2\pi)^6 p_E^6} \ =\ \frac 1{64\pi^3} \ln \frac \Lambda \mu\ .
  \ee
The graph in Fig. 1b gives the counterterm  $\propto D^3$. We obtain
  \be
\label{DLD3}
\Delta {\cal L}_{\rm eff}^{D^3} \ =\ - \frac {2g^3c_VL}{3} {\rm Tr} \{ D_{ij}  D^{jm} D^i_{\ m} \}\ . 
 \ee
Adding Eqs. (\ref{DLpD2}), (\ref{DLD3}) to the corresponding terms in the free lagrangian (\ref{CompAct}) 
and performing a proper
scale redefinition of the field $D$ gives us  the renormalization of the effective charge,
 \be
\label{renorm}
g(\mu) \ =\ g_0 \left( 1 - \frac {g_0^2 L c_V}3 \right).
 \ee
The sign of the correction corresponds to the Landau zero situation, like in the pure SYM case, but the value of the 
coefficient in (\ref{renorm})
turned out to be 20 times smaller than that  for pure HD SYM.

\section*{Acknowledgements}

E.I. thanks the SUBATECH, Nantes, and A.S. thanks the BLTP JINR, Dubna,  
for the kind hospitality extended to them in the course of this study. 
E.I. acknowledges 
a partial support from RFBR grants, projects  No
03-02-17440 and No 02-02-04002, NATO grant PST.GLG.980302, 
the DFG grant No.436 RUS 113/669-02, and a grant of the
Heisenberg-Landau program.

\section*{Appendix A}
\renewcommand{\theequation}{A\arabic{equation}}
\setcounter{equation}0
 We prove here Eq. (\ref{delSn}).

Introduce the notation
 \be
\label{defIJ}
I_{nm} = {\rm Tr}\, q^{+A} (\nabla^{--})^n  (\nabla^{++})^m q^+_A\ , \quad
J_{nm} = {\rm Tr}\, q^{+A} (\nabla^{++})^n  (\nabla^{--})^m q^+_A\ .
 \ee
By a multiple use of commutation relation (\ref{D00}), one can derive
 \be
\label{relIJ}
J_{nn} &=& I_{nn} + n^2 I_{n-1,n-1}\, , \nonumber \\
I_{n+1,n-1} &=& J_{n-1,n+1} + (n^2-1) J_{n-2,n}  \, , \nonumber \\
I_{n,n-1} &=& J_{n-1,n} = {\cal L}_n\, .
 \ee
Calculate now the conformal variation of $S_{n+1}\,$, using the variations 
of the harmonic derivatives (Eq. \p{NablaTran}),
of the hypermultiplet $q^+$ (Eq.(\ref{qcftran})) and of the measure $\mu$ (Eq. \p{meastran}), 
and taking into account the properties
 \be
\label{Dlam}
D^{\pm\pm} \lambda^{\mp\mp} = \lambda, \ \ \ \ \ D^{\pm\pm} \lambda = 2D^{\pm\pm} (\eta^-\theta^+) = 
 2\lambda^{\pm\pm}, \ \ \ \ \ 
D^{\pm\pm} \lambda^{\pm\pm} = 0 
 \ee
(recall that $\lambda = \eta^-\theta^+ + \eta^+\theta^-\,$, $\lambda^{\pm\pm} 
= \eta^\pm \theta^\pm$). Defining 
\be
\delta_\eta S_{n+1} = \int \mu\, \Delta_\eta {\cal L}_{n+1}\,, \quad \Delta_\eta {\cal L}_{n+1} = 
\delta_\eta {\cal L}_{n+1} + i(3\eta^+\theta^- + \eta^-\theta^+){\cal L}_{n+1}\,,
\ee
where the second term in $\Delta_\eta {\cal L}_{n+1}$ comes from the tarnsformation 
of the measure $\mu\,$, we obtain, modulo a total harmonic derivative, 
\be
\Delta_\eta {\cal L}_{n+1} = i(1-2n) ( \lambda {\cal L}_{n+1} + Q_n)\, ,
 \ee
where
 \be
\lb{Qndef}
Q_n = (n+1) \lambda^{--} J_{nn} + n\lambda^{++} I_{n+1,n-1}   + n(n+1) \lambda {\cal L}_n \, .
 \ee 
When deriving (\ref{Qndef}), we used also the first of the relations (\ref{relIJ}). 

Now we will give an inductive proof that $Q_n$ is reduced to a total derivative and so 
does not contribute to $\delta_\eta S_{n+1}\,$. Pulling out the harmonic derivatives 
in the first and the second terms in (\ref{Qndef}) from the right to the left and throwing away 
the total harmonic derivatives in the process, we obtain (up to total derivatives)
 \be
\label{perekrut}
\lambda^{--} J_{nn} &=& - {\rm Tr}\, q^{+A} (D^{--})^n  (D^{++})^n \lambda^{--} q^+_A\ , \nonumber \\
\lambda^{++} I_{n+1,n-1}  &=&  - {\rm Tr}\,q^{+A} (D^{++})^{n-1}  
(D^{--})^{n+1} \lambda^{++} q^+_A\ .
 \ee
Using Eqs. (\ref{Dlam}) and (\ref{relIJ}), one can derive the recurrency relations
 \be
\label{recur}
\lambda^{--} J_{nn} &=& -\lambda^{--} \frac {n^2}2 J_{n-1,n-1} 
- \lambda \frac n2 [{\cal L}_n +n(n-1){\cal L}_{n-1}] \nn \\
&& -\, \frac {n(n-1)}2 \lambda^{++} I_{n,n-2}  
= - \frac n2 (\lambda {\cal L}_n + Q_{n-1})\, , \nonumber \\
\lambda^{++} I_{n+1,n-1}  &=&   - 
\frac {n(n+1)}2 \lambda^{--} J_{n-1,n-1}
 - \lambda \frac {n+1}2 [{\cal L}_n +n(n-1){\cal L}_{n-1}] \nn \\
&& - \frac {n^2-1}2 \lambda^{++} I_{n,n-2} 
=  - \frac {n+1}2 (\lambda {\cal L}_n + Q_{n-1})\, ,
 \ee
where as before the equality is understood in a weak sense, i.e. modulo total harmonic derivatives.  
With the inductive assumption $Q_{n-1} = 0$, we obtain
 \be 
 \label{IJresult}
n \lambda^{++} I_{n+1,n-1} = (n+1) \lambda^{--} J_{nn} = - \frac {n(n+1)}2 \lambda {\cal L}_n\,,
 \ee
which implies the sought result $Q_n = 0\,$ and thus proves (\ref{delSn}).
\setcounter{equation}{0}

\section*{Appendix B}
\renewcommand{\theequation}{B\arabic{equation}}
In this Appendix we show how some actions from the set $S_n$ defined in \p{qnab--q}, \p{Sn} look 
in the analytic superspace and, just to give a feeling of what their component structure is,    
present the component form of the free part of the action $S_2\,$.  

We start with $S_1\,$. Using the relation \p{RelMea} between the integration measures of the full $6D$ 
HSS and its analytic subspace, we obtain 
\be
S_1 = 2 \mbox{Tr} \int d^6x_A du d^4\theta^+ q^{+ A} F^{++} q^+_A\,, \quad 
F^{++} = (D^{+})^4 V^{--}\,.\label{S11}
\ee
We see that this scale-invariant $6D$ action has no free part and contains 
some non-minimal interactions between hypermultiplet and gauge multiplet. 
The component structure of $F^{++}$ was established in \cite{ISZ}, 
so it is in fact straightforward to find the component form of \p{S11}.  

The actions $S_n$ with $n>1$ all contains nontrivial higher-derivative free parts. 
Using the formula \cite{ISZ}
\be
\frac12 (D^+)^4(D^{--})^2 \Phi = \Box \Phi\,,  \label{Box1}
\ee
which is valid for any analytic superfield $\Phi$, it is easy to find the analytic superspace 
form of the action $S_2$ with all gauge fields discarded
\be
S_2^{free} = \mbox{Tr} \int \mu q^{+A}(D^{--})^2D^{++} q^+_A = 
2 \mbox{Tr} \int d^6x_A du d^4\theta^+ q^{+ A} \Box D^{++}q^+_A\,.\label{S2free}
\ee
In the case with the gauge superfields retained, the corresponding analytic superspace 
action includes a non-trivial covariantization of the box $\Box$. 
For $4D$ case the explicit form of the covariant box can be found e.g. in \cite{Backgr}.

Our last example is $S_3\,$. Using an analog of the formula \p{Box1} with $(D^{--})^3$, it 
is easy to show that 
\be
S_3^{free} = 6 \mbox{Tr} \int d^6x_A du d^4\theta^+ q^{+ A} 
\left(\Box \partial^{--} + i\partial_{+a}\partial_{+b}\partial^{ab}\right) (D^{++})^2q^+_A\,.
\ee 

Finally, we present the component form of $S_2^{free}\,$. Using the $\theta$ expansion
\be
\label{qplA}
q^{+ A} = f^{+ A} + \theta^{+a}\psi^A_a + \frac12 \theta^+\gamma^M\theta^+ F^{-A}_M + 
(\theta^+)^3_a\chi^{-2 a} + (\theta^+)^4 F^{-3 A}\,, 
\ee
where \cite{ISZ} $(\theta^+)^3_a = \frac16\epsilon_{cbda}\theta^{+c} \theta^{+b} \theta^{+d}$, 
$ (\theta^+)^4 = -\frac{1}{24}\epsilon_{abcd}\theta^{+a} \theta^{+b} \theta^{+c} \theta^{+d}$,
it is easy to do the $\int d^4\theta^-$ integral in \p{S2free} over $\theta^{+a}$ to obtain 
\be
S_2^{free} &\sim & \mbox{Tr} \int d^6 x du \left(2f^{+A}\partial^{++} \Box F^{-3}_A -4\mbox{i} 
f^{+A}\Box \partial_M F^{- M}_A - 2F^{-A}_M \partial^{++}\Box F^{-M}_A \right. \nn \\
&& \left. +\, \mbox{i} \psi^A\tilde{\gamma}^M\Box \partial_M\psi_A 
- 2\psi^A_a\partial^{++}\Box \chi^{-2 a}_A \right). 
\ee  
 
We see that all the components in (\ref{qplA})  are now propagating and that we cannot suppress 
their harmonic dependence. 
Expansion over harmonics gives  an infinite set of propagating fields with growing isospins.

\end{document}